\documentclass{article}
\usepackage{amsfonts}

\usepackage{graphicx}
\usepackage{amsmath}

\newtheorem{theorem}{Theorem}

\newtheorem{corollary}[theorem]{Corollary}

\newtheorem{definition}[theorem]{Definition}

\newtheorem{lemma}[theorem]{Lemma}

\newtheorem{proposition}[theorem]{Proposition}
\newtheorem{remark}[theorem]{Remark}

\newenvironment{proof}[1][Proof]{\textbf{#1.} }{\ \rule{0.5em}{0.5em}}

\begin{document}

\title{Nambu-Poisson manifolds and associated n-ary Lie algebroids}
\author{Jos\'{e} A. Vallejo\thanks{%
Work supported by the Spanish Ministerio de Educaci\'{o}n y Cultura, FPI
grant N${{}^{o}}$PB-97 1386} \thanks{%
e-mail address: Jose.A.Vallejo@uv.es} \\
Departament de Geometria i Topologia\\
Universitat de Val\`{e}ncia\\
46100 Burjassot (Valencia), Spain}
\maketitle

\begin{abstract}
We introduce an n-ary Lie algebroid canonically associated to a
Nambu-Poisson manifold. We also prove that every Nambu-Poisson bracket
defined on functions is induced by some differential operator on the
exterior algebra, and characterize such operators. Some physical examples
are presented.
\end{abstract}

\section{Introduction}

In the seventies, trying to describe the simultaneous classical dynamics of
three particles as a previous step towards a quantum statistics for the
quark model, Y. Nambu introduced a generalization of Poisson brackets
formalism that today bears his name (see [Nam 73]). The field underwent a
revitalization with the work of L. Takhtajan in the nineties ([Tak 94]),
which showed the algebraic setting underlying Nambu ideas and introduced an
analog of the Jacobi Identity, the Fundamental Identity, allowing the
connection with the theory of n-ary Lie algebras developed by Filippov and
others (see [Fil 85] and [Han-Wac 95]). In [AIP 97], further references
about the development of this subject can be found, along with an
alternative generalization of Poisson structures.

Recently, much work has been done in this area, showing an interesting
algebraic structure (see, for instance, [Ale-Guh 96], [Gau 96], [Gau 98],
[Mic-Vin 96], [MVV 98], [Nak 98], [Vin-Vin 98] and references therein). This
is very interesting both from a mathematical and a physical point of view:
an algebraic formulation not only provides a more concise and elegant
framework, but also can be the source for new insights. In this direction,
the notion of Leibniz algebroid was introduced ([Dal-Tak 97], [ILMP 99]) as
a kind of analogue of the Lie algebroid associated to a Poisson manifold.
That concept of a Lie algebroid, has proved to be very useful in the study
of Poisson manifolds and dynamics (for a sample, see [Lib 96], [Cou 94],
[Wei 98], [Gra-Urb 97]), and recently has been considered for the
quantization of Poisson algebras ([Lan-Ram 00]). It is to be expected the
same in the Nambu-Poisson case (n-ary Poisson algebras), with the appropiate
generalizations.

Thus, the main motivation for this work comes from the question: to what
extent the constructions and properties of the Lie algebroids associated to
Poisson manifolds, including quantization, carry over to the n-ary case?
This paper deals with a very first step towards the answer, namely, the
study of the canonical relation between Lie algebroids and Poisson manifolds
in the n-ary case. For this purpose, one could consider Leibniz algebroids;
however, this concept (although very interesting in itself) is not a genuine
generalization of that of a Lie algebroid: it relies upon an algebraic
construction called Leibniz (or Loday) algebra (see [Lod 93]), which is a
non-commutative version of a Lie algebra. In [Gra-Mar 00] the notion of an
n-ary Lie algebroid (called by the authors Filippov algebroid) has been
introduced, and it seems to fit better to the aforementioned question.

In this notes, we construct an n-ary Lie algebroid canonically associated to
a Nambu-Poisson manifold \`{a} la Koszul (see [Kos 85]), that is to say, by
using differential operators on the (graded) exterior algebra of
differential forms. An interesting advantage of this approach, is that it
allows for the possibility of characterizing the operators which generate
the brackets under consideration, this characterization being in terms of
the commutator of the operator with the exterior differential. For the case
of Poisson brackets and its extension as a graded brackets to the whole
exterior algebra, such study was made in [BMS 97], where these operators
were called Jacobi operators (adopting the terminology from [Gra 92]), and
essentially the same techniques will be used here to show that every
Nambu-Poisson bracket coincides with the bracket induced on functions by a
certain differential operator on the exterior algebra, giving its explicit
form in terms of the Nambu-Poisson multivector. In the last section, some
physical examples are described.

\textbf{Acknowledgments:} The author wants to express his gratitude to J. V.
Beltr\'{a}n, J. Grabowski, J. A. de Azc\'{a}rraga and specially J. Monterde,
for very useful discussions and comments.

\section{Basic definitions and results}

Our main tool in the study of the n-ary generalizations of Poisson manifolds
and Lie algebroids, will be the theory of differential operators on the
exterior algebra of a manifold, so we collect here the basics of this theory.

Let $Hom_{\mathbb{R}}(\Omega (M))$ be the space of $\mathbb{R}-$%
homomorphisms of the exterior algebra $F:\Omega (M)\longrightarrow \Omega
(M) $. We say that $F$ has $\mathbb{Z}$-degree $\left| F\right| $ (sometimes 
$F$ as exponent) if $F$ maps $p-$forms on $(p+\left| F\right| )-$forms, then
we write $F\in Hom_{\mathbb{R}}^{\left| F\right| }(\Omega (M))$ (or $F\in
Hom_{\mathbb{R}}^{F}(\Omega (M))$ ).

On the space $Hom_{\mathbb{R}}(\Omega (M))$ we introduce a bracket $[.,.]$
(called commutator) by means of 
\begin{equation*}
\lbrack F,G]=F\circ G-(-1)^{FG}G\circ F
\end{equation*}
and it is easy to prove that this bracket turns $(Hom_{\mathbb{R}}(\Omega
(M))$ $,[.,.])$ into a graded Lie algebra.

A differential operator on $\Omega (M)$ of degree $p$ and order $q$, is a
homomorphism $D\in Hom_{\mathbb{R}}^{p}(\Omega (M))$ such that 
\begin{equation*}
\lbrack \lbrack ...[[D,\mu _{a_{0}}],\mu _{a_{1}}],...],\mu _{a_{q}}]=0
\end{equation*}
for all $\mu _{a_{i}},i\in \{0,...,q\}$, where $a_{i}\in \Omega (M)$ and $%
\mu _{a_{i}}$ denotes the homomorphism multiplication by $a_{i},\mu
_{a_{i}}(b)=a_{i}\wedge b$, which has order 0 and degree $\left|
a_{i}\right| $, and so will be often denoted simply $a_{i}$. The space of
such operators is denoted $\mathcal{D}_{q}^{p}$; examples are the insertion
operator $i_{P}\in \mathcal{D}_{p}^{-p}$ where $P\in \Gamma (\Lambda ^{p}TM)$
is a $p-$multivector, and the generalized Lie derivative $\mathcal{L}%
_{P}=[i_{P},d]\in \mathcal{D}_{p}^{-(p-1)}$, $d$ being the exterior
differential.

A useful result states that 
\begin{equation}
\lbrack \mathcal{D}_{q}^{p},\mathcal{D}_{q^{\prime }}^{p^{\prime }}]\subset 
\mathcal{D}_{q+q^{\prime }-1}^{p+p^{\prime }}.  \label{eq1}
\end{equation}

Also, we have that any operator $D\in \mathcal{D}_{q}^{-q}$ has the form $%
i_{A}$, for some $A\in \Gamma (\Lambda ^{q}TM)$. This is a consequence of (%
\ref{eq1}) and the fact that any operator is determined by its action on
forms with degree equal or less than the order of the operator.

The order defines a filtration on the space of all differential operators.
Indeed, given $p,k\in \mathbb{Z}$ with $p>k$, if $D$ is an operator of order 
$\leq k$ , then it is also of order $\leq p$, that is to say: 
\begin{equation*}
\{0\}=\mathcal{D}_{-1}(M)\subset \mathcal{D}_{0}(M)\subset ...\subset 
\mathcal{D}_{k}(M)\subset \mathcal{D}_{k+1}(M)\subset ...
\end{equation*}

\begin{definition}
Given $k\in \mathbb{Z}$, we will call symbol of order $k$ of $M$ the $\Omega
(M)$-module 
\begin{equation*}
Symb_{k}(M)=\frac{\mathcal{D}_{k}(M)}{\mathcal{D}_{k-i}(M)},
\end{equation*}
and the space of symbols of the exterior algebra $\Omega (M)$ is defined as
the graded $\Omega (M)$-module 
\begin{equation*}
Symb(M)=\underset{k\in \mathbb{Z}}{\cup }Symb_{k}(M).
\end{equation*}
Note that, if $D$ is a differential operator of order $\leq k$, $D\in 
\mathcal{D}_{k}(M)$, then we can speak about the symbol of order $i$ of the
operator $D$, that is to say, 
\begin{equation*}
Symb_{i}(D)\in \frac{\mathcal{D}_{k}(M)}{\mathcal{D}_{k-1}(M)},\text{ \ \ }%
i\leq k.
\end{equation*}
\end{definition}

We will need some terminology from the theory of Nambu-Poisson manifolds.

A Nambu-Poisson bracket on a manifold $M$, is an $n$-linear mapping $%
\{.,...,.\}:C^{\infty }(M)\times ...\times C^{\infty }(M)\longrightarrow
C^{\infty }(M)$, satisfying:

\begin{enumerate}
\item  Skew-symmetry. For all $f_{i}\in C^{\infty }(M),(1\leq i\leq n)$ and $%
\sigma \in S_{n}$ ($S_{n}$ is the symmetric group of order $n$) 
\begin{equation*}
\{f_{1},...,f_{n}\}=(-1)^{\varepsilon (\sigma )}\{f_{\sigma
(1)},...,f_{\sigma (n)}\}
\end{equation*}

\item  Leibniz rule. For all $f_{i},g\in C^{\infty }(M),(1\leq i\leq n)$ 
\begin{equation*}
\{f_{1}g,f_{2},...,f_{n}\}=f_{1}\{g,f_{2},...,f_{n}\}+g%
\{f_{1},f_{2},...,f_{n}\}
\end{equation*}

\item  Fundamental Identity. For all $f_{i},g_{j}\in C^{\infty }(M),(1\leq
i\leq n-1,1\leq j\leq n)$%
\begin{equation*}
\{f_{1},f_{2},...,f_{n-1},\{g_{1},...,g_{n}\}\}=\sum\limits_{j=1}^{n}%
\{g_{1},...,\{f_{1},...,f_{n-1},g_{j}\},...,g_{n}\}
\end{equation*}
\end{enumerate}

If $\{.,...,.\}$ is a Nambu-Poisson bracket, it has associated an $n$%
-multivector $P\in \Gamma (\Lambda ^{n}TM)$ through 
\begin{equation*}
P(df_{1}\wedge ...\wedge df_{n})=\{f_{1},...,f_{n}\},
\end{equation*}
which is called the Nambu-Poisson multivector.

A Nambu-Poisson manifold is a pair $(M,\{.,...,.\})$ where $\{.,...,.\}$ is
a Nambu-Poisson bracket (also can be denoted $(M,P)$).

A Lie (or Filippov) $n$-algebra structure on a vector space $V$, is an $n$%
-linear skew-symmetric bracket $[.,...,.]$ satisfying the generalized Jacobi
identity, also called Fundamental Identity: 
\begin{equation*}
\lbrack
v_{1},v_{2},...,v_{n-1},[u_{1},...,u_{n}]]=\sum%
\limits_{j=1}^{n}[u_{1},...,[v_{1},...,v_{n-1},u_{j}],...,u_{n}]
\end{equation*}
for all $v_{i},u_{j}\in V,(1\leq i\leq n-1,1\leq j\leq n)$.

A Lie (or Filippov) $n$-algebroid is a vector bundle $p:E\longrightarrow M$
equipped with an $n$-bracket $[.,...,.]$ on sections of $E$ ($V=\Gamma (E)$)
and a vector bundle morphism $q:\Lambda ^{n-1}E\longrightarrow TM$ over the
identity on $M$, called the anchor map, such that:

\begin{enumerate}
\item  The induced morphism on sections $q:\Gamma (\Lambda
^{n-1}E)\longrightarrow \Gamma (TM)$ satisfies the following relation with
respect to the bracket of vector fields 
\begin{multline*}
\lbrack q(X_{1}\wedge ...\wedge X_{n-1}),q(Y_{1}\wedge ...\wedge Y_{n-1})]=
\\
\sum\limits_{i=1}^{n-1}q(Y_{1}\wedge ...[X_{1},...,X_{n-1},Y_{i}]\wedge
...\wedge Y_{n-1})
\end{multline*}

\item  and 
\begin{equation*}
\lbrack X_{1},...,X_{n-1},fY]=f[X_{1},...,X_{n-1},Y]+q(X_{1}\wedge ...\wedge
X_{n-1})(f)Y
\end{equation*}
for all $X_{1},...,X_{n-1},Y\in \Gamma (E)$ and $f\in C^{\infty }(M)$.
\end{enumerate}

We will call these structures n-Lie brackets and algebroids, respectively,
for short.

\section{n-Lie brackets on 1-forms induced by differential operators}

In [Kos 85], Koszul introduces the following notation: for $D\in \mathcal{D}%
_{n}$ (of any degree) 
\begin{equation}
\Phi _{D}^{n}(a_{1},...,a_{n})=[[...[D,a_{1}],...],a_{n}](1).  \label{eq2}
\end{equation}

In fact, he considers that expression for $D\in \mathcal{D}_{2}^{-1}$ such
that $D(1)=0$, and uses it in order to define a bracket on $\Omega (M)$ as 
\begin{eqnarray}
\lbrack {}\![a,b]\!]_{D} &=&(-1)^{a}\Phi _{D}^{2}(a,b)=(-1)^{a}[[D,a],b](1)
\label{eq3} \\
&=&(-1)^{a}D(ab)-(-1)^{a}D(a)b-(-1)^{a(D+1)}aD(b).  \notag
\end{eqnarray}

Koszul also studies under what conditions the bracket $[\![.,.]\!]_{D}$ is a
graded Lie one, and obtains the necessary and sufficient condition 
\begin{equation*}
D^{2}\in \mathcal{D}_{2},
\end{equation*}
when a priori $D^{2\text{ }}$lies in $\mathcal{D}_{3}$. This idea can be
followed on to construct new kinds of brackets from a differential operator.

\begin{remark}
In the following, we will consider only operators such that $D(1)=0$.
\end{remark}

By a direct calculation involving only Jacobi identity and the skew-symmetry
for the commutator $[.,.]$, it can be shown that, for any $D\in \mathcal{D}%
_{n}^{D}$%
\begin{gather}
\Phi _{D}^{n}(a_{1},...,a_{n-1},\Phi _{D}^{n}(b_{1},...,b_{n}))=  \label{eq4}
\\
(-1)^{(D+\sum\limits_{i=1}^{n-1}a_{i})D}\Phi _{D}^{n}(\Phi
_{D}^{n}(a_{1},...,a_{n-1},b_{1}),b_{2},...,b_{n})  \notag \\
+(-1)^{(D+\sum\limits_{i=1}^{n-1}a_{i})(D+b_{1})}\Phi _{D}^{n}(b_{1},\Phi
_{D}^{n}(a_{1},...,a_{n-1},b_{2}),b_{3},...,b_{n})  \notag \\
...  \notag \\
+(-1)^{^{(D+\sum\limits_{i=1}^{n-1}a_{i})(D+\sum\limits_{j=1}^{n-1}b_{j})}}%
\Phi _{D}^{n}(b_{1},...,b_{n-1},\Phi _{D}^{n}(a_{1},...,a_{n-1},b_{n})) 
\notag \\
+[[...[[[[...[D,a_{1}],...],...],a_{n-1}],D],b_{1}],...],b_{n}](1).  \notag
\end{gather}
If we want to induce a n-ary bracket on 1-forms, we need an operator of type 
$D\in \mathcal{D}_{n}^{-(n-1)}(\Omega (M))$ (so acting on $n$ 1-forms gives
a 1-form, recall (\ref{eq1})). In this case, with $a_{i},b_{j}\in \Omega
^{1}(M)$, (\ref{eq4}) reduces to 
\begin{gather*}
\Phi _{D}^{n}(a_{1},...,a_{n-1},\Phi _{D}^{n}(b_{1},...,b_{n}))= \\
(-1)^{(D+n-1)D}\Phi _{D}^{n}(\Phi
_{D}^{n}(a_{1},...,a_{n-1},b_{1}),b_{2},...,b_{n}) \\
+(-1)^{(D+n-1)(D+1)}\Phi _{D}^{n}(b_{1},\Phi
_{D}^{n}(a_{1},...,a_{n-1},b_{2}),b_{3},...,b_{n}) \\
... \\
+(-1)^{^{(D+n-1)(D+n-1)}}\Phi _{D}^{n}(b_{1},...,b_{n-1},\Phi
_{D}^{n}(a_{1},...,a_{n-1},b_{n})) \\
+[[...[[[[...[D,a_{1}],...],a_{n-1}],D],b_{1}],...],b_{n}](1).
\end{gather*}
and, as the operator degree is $\left| D\right| =-(n-1)$, to 
\begin{gather}
\Phi _{D}^{n}(a_{1},...,a_{n-1},\Phi _{D}^{n}(b_{1},...,b_{n}))=  \label{eq5}
\\
\Phi _{D}^{n}(\Phi _{D}^{n}(a_{1},...,a_{n-1},b_{1}),b_{2},...,b_{n})  \notag
\\
+\Phi _{D}^{n}(b_{1},\Phi _{D}^{n}(a_{1},...,a_{n-1},b_{2}),b_{3},...,b_{n})
\notag \\
...  \notag \\
+\Phi _{D}^{n}(b_{1},...,b_{n-1},\Phi _{D}^{n}(a_{1},...,a_{n-1},b_{n})) 
\notag \\
+[[...[[[[...[D,a_{1}],...],a_{n-1}],D],b_{1}],...],b_{n}](1).  \notag
\end{gather}

Taking into account that (for odd degree operators) $[[D,a],D]=\frac{1}{2}[%
D^{2},a]$, we see that this expression generalizes the one given by Koszul
(in lemma 1.5 of [Kos 85]) for the Jacobi Identity when only 1-forms are
considered: 
\begin{gather*}
\Phi _{D}^{2}(a_{1},\Phi _{D}^{2}(a_{2},a_{3}))= \\
\Phi _{D}^{2}(\Phi _{D}^{2}(a_{1},a_{2}),a_{3})+\Phi _{D}^{2}(a_{2},\Phi
_{D}^{2}(a_{1},a_{3}))+\frac{1}{2}\Phi _{D^{2}}^{3}(a_{1},a_{2},a_{3}).
\end{gather*}

Thus, we are led to consider the following definition for a n-ary bracket on
1-forms, induced by an operator $D\in \mathcal{D}_{n}^{-(n-1)}.$

\begin{definition}
For $D\in \mathcal{D}_{n}^{-(n-1)}$ and $a_{i}\in \Omega ^{1}(M)$, $i\in
\{i,...,n\},$%
\begin{equation}
\lbrack \![a_{1},...,a_{n}]\!]_{D}=\Phi
_{D}^{n}(a_{1},...,a_{n})=[[...[D,a_{1}],...],a_{n}](1).  \label{eq6}
\end{equation}
\end{definition}

\begin{proposition}
Let $[\![.,...,.]\!]_{D}$ the bracket on 1-forms induced by an operator $%
D\in \mathcal{D}_{n}^{-(n-1)}$. Then, it has the following properties:

(i) $\mathbb{R}$-linearity on each argument

(ii) skew-symmetry.
\end{proposition}

\begin{proof}
(i) Is a direct consequence of the corresponding property for the bracket on 
$Hom_{\mathbb{R}}(\Omega (M))$.

(ii) Here we use that, for any $F\in Hom_{\mathbb{R}}(\Omega (M))$, and $%
a\in \Omega ^{\left| a\right| }(M),b\in \Omega ^{\left| b\right| }(M),$%
\begin{equation*}
\lbrack \lbrack F,a],b]=(-1)^{ab}[[F,b],a]
\end{equation*}
and the statement follows from a straightforward computation.
\end{proof}

In order to construct an n-Lie bracket on 1-forms, we need a Fundamental
Identity. A glance at (\ref{eq5}) tells us we are almost done, it suffices
to give a condition on $D$ similar to that of Koszul for the binary case.

\begin{proposition}
Let\ $D\in \mathcal{D}_{n}^{-(n-1)}$ be a differential operator. Then, $%
[\![.,...,.]\!]_{D}$ induces an n-Lie algebra structure on $\Omega ^{1}(M)$
if and only if, for any $a_{1},...,a_{n-1}\in \Omega ^{1}(M)$, 
\begin{equation*}
\lbrack \lbrack \lbrack ...[D,a_{1}],...],a_{n-1}],D]\in \mathcal{D}_{n-1}.
\end{equation*}
\end{proposition}

\begin{proof}
Just observe that the condition above kills the last term in (\ref{eq5}).
\end{proof}

\section{n-Lie algebroids}

In this section, we will construct an n-Lie algebroid on the space of $1-$%
forms. Given an operator $D\in \mathcal{D}_{n}^{-(n-1)}$, we consider an
extension of the bracket $[\![.,...,.]\!]_{D}:\Omega ^{1}(M)\times .\overset{%
n)}{.}.\times \Omega ^{1}(M)\longrightarrow \Omega ^{1}(M)$ to another one
defined in $\Omega ^{1}(M)\times .\overset{n-1)}{.}.\times \Omega
^{1}(M)\times \Omega (M)$, that is, where the last argument is a
differential form of any degree, with the same formula: 
\begin{equation*}
\lbrack \![a_{1},...,a_{n-1},b]\!]_{D}=\Phi _{D}^{n}(a_{1},...,a_{n-1},b),
\end{equation*}
$\forall a_{i}\in \Omega ^{1}(M),i\in \{1,...,n-1\},b\in \Omega (M)$.
Although this bracket loses the skew-symmetry property, under the condition
on $D$ given in Proposition 2 it retains the Fundamental Identity, in the
sense that (recall (\ref{eq4})) 
\begin{gather*}
\Phi _{D}^{n}(a_{1},...,a_{n-1},\Phi _{D}^{n}(b_{1},...,b_{n-1},f))= \\
\Phi _{D}^{n}(\Phi _{D}^{n}(a_{1},...,a_{n-1},b_{1}),b_{2},...,b_{n-1},f)+ \\
... \\
+\Phi _{D}^{n}(b_{1},...,b_{n-1},\Phi _{D}^{n}(a_{1},...,a_{n-1},f)),
\end{gather*}
when $f\in C^{\infty }(M),a_{i},b_{i}\in \Omega ^{1}(M),i\in \{1,...,n-1\}$.
This will enable us to construct an n-Lie (or Filippov) algebroid associated
to $D$.

Indeed, the Leibniz property guarantees that $[\![a_{1},...,a_{n-1},.]%
\!]_{D} $ acts as a derivation on $C^{\infty }(M)$, that is, it belongs to $%
\Gamma (TM)$: for any $f,g\in C^{\infty }(M),a_{i}\in \Omega ^{1}(M),i\in
\{1,...,n-1\}$, 
\begin{gather*}
\lbrack \![a_{1},...,a_{n-1},fg]\!]_{D}= \\
=f[\![a_{1},...,a_{n-1},g]\!]_{D}+[\![a_{1},...,a_{n-1},f]\!]_{D}g.
\end{gather*}

Consider now the cotangent bundle $T^{\ast }M\overset{\pi }{\longrightarrow }%
M$. On the sections $\Omega ^{n-1}(M)=\Gamma (\Lambda ^{n-1}(T^{\ast }M))$,
we define the anchor map as 
\begin{align*}
q& :\Omega ^{n-1}(M)\longrightarrow \Gamma (TM) \\
a_{1}\wedge ...\wedge a_{n-1}& \mapsto q(a_{1}\wedge ...\wedge
a_{n-1})=[\![a_{1},...,a_{n-1},.]\!]_{D}
\end{align*}
and extend it by linearity. Let us check that the definition makes sense:
the observation above tells us that for $f,g\in C^{\infty }(M),$ 
\begin{equation*}
q(a_{1}\wedge ...\wedge a_{n-1})(fg)=fq(a_{1}\wedge ...\wedge
a_{n-1})(g)+gq(a_{1}\wedge ...\wedge a_{n-1})(f),
\end{equation*}
and so $q$ takes values in the right space. Note also the $C^{\infty }(M)-$%
linearity of the anchor map, obtained as a consequence of the Leibniz
property for the bracket on $Hom_{\mathbb{R}}(\Omega (M))$ and the degree of 
$D$: if $f,g\in C^{\infty }(M),a_{i}\in \Omega ^{1}(M),i\in \{1,...,n-1\}$, 
\begin{gather*}
q(a_{1}\wedge ...\wedge fa_{i}\wedge ...\wedge
a_{n-1})(g)=[\![a_{1},...,fa_{i},...,a_{n-1},g]\!]_{D}= \\
=[[...[[...[D,a_{i}],...],fa_{i}],...],g](1)= \\
=f[[...[[...[D,a_{1}],...],a_{i}],...],g](1)+(-1)^{a_{i}(n-i-1)}[[...[[...[D,a_{1}],...],f],...],g](a_{i})=
\\
=fq(a_{1}\wedge ...\wedge a_{i}\wedge ...\wedge a_{n-1})(g)
\end{gather*}
(the term with $f,g$ inside the brackets vanishes because $\left| D\right|
=-(n-1)$).

Let us now verify the conditions of the n-Lie algebroid definition. One of
these is nothing more but the property (ii) of Proposition 1: 
\begin{gather*}
\lbrack
\![a_{1},...,a_{n-1},fb]\!]_{D}=f[\![a_{1},...,a_{n-1},b]\!]_{D}+[%
\![a_{1},...,a_{n-1},f]\!]_{D}b= \\
=f[\![a_{1},...,a_{n-1},b]\!]_{D}+q(a_{1}\wedge ...\wedge a_{n-1})(f)b
\end{gather*}
and the other one is equivalent to the Fundamental Identity. On the one hand
we have 
\begin{gather*}
\lbrack q(a_{1}\wedge ...\wedge a_{n-1}),q(b_{1}\wedge ...\wedge
b_{n-1})](f)= \\
=q(a_{1}\wedge ...\wedge a_{n-1})\!\
[\![b_{1},...,b_{n-1},f]\!]_{D}-q(b_{1}\wedge ...\wedge
b_{n-1})[\![a_{1},...,a_{n-1},f]\!]_{D}= \\
=[\![a_{1},...,a_{n-1},[\![b_{1},...,b_{n-1},f]\!]_{D}]\!]_{D}-[%
\![b_{1},...,b_{n-1},[\![a_{1},...,a_{n-1},f]\!]_{D}]\!]_{D}
\end{gather*}
and on the other 
\begin{gather*}
\sum\limits_{i=1}^{n-1}q(b_{1}\wedge ...\wedge \lbrack
\![a_{1},...,a_{n-1},b_{i}]\!]_{D}\wedge ...\wedge b_{n-1})(f)= \\
\sum\limits_{i=1}^{n-1}[\![b_{1},...,[\![a_{1},...,a_{n-1},b_{i}]%
\!]_{D},....b_{n-1},f]\!]_{D}.
\end{gather*}

The Fundamental Identity equates these two expressions, so 
\begin{gather*}
\lbrack q(a_{1}\wedge ...\wedge a_{n-1}),q(b_{1}\wedge ...\wedge
b_{n-1})](f)= \\
\sum\limits_{i=1}^{n-1}q(b_{1}\wedge ...\wedge \lbrack
\![a_{1},...,a_{n-1},b_{i}]\!]_{D}\wedge ...\wedge b_{n-1})(f).
\end{gather*}

We summarize all this in the following result.

\begin{theorem}
Let $D\in \mathcal{D}_{n}^{-(n-1)}$ be such that $%
[[[...[D,a_{1}],...],a_{n-1}],D]\in \mathcal{D}_{n-1}$, $\forall a_{i}\in
\Omega ^{1}(M),i\in \{1,...,n-1\}$. Then $\left( T^{\ast }M\overset{\pi }{%
\longrightarrow }M,[\![.,...,.]\!]_{D},q\right) $ is an n-Lie algebroid.
\end{theorem}

\begin{remark}
Given an operator $D$ as in the previous Proposition, any isomorphism $%
L:T^{\ast }M\longrightarrow TM$ (for example, the canonical ones associated
to riemannian, Poisson or symplectic manifolds) induces the corresponding
n-Lie algebroid on $\left( TM\overset{\pi }{\longrightarrow }%
M,[\![.,...,.]\!]_{\tilde{D}},q\right) $, where $\tilde{D}=\tilde{L}\circ
D\circ \tilde{L}^{-1}$ and $\tilde{L}$ is the extension of $L$ as a
homomorphism of exterior algebras.
\end{remark}

\section{The canonical n-Lie algebroid associated to a Nambu-Poisson manifold%
}

Our goal in this section is to construct a basic example of n-Lie algebroid,
and we shall obtain a result similar to that for Poisson manifolds (namely,
that each Poisson manifold has a canonical Lie algebroid structure
associated to it) for the n-ary case, i.e: each n-Poisson (Nambu-Poisson)
manifold has a canonically associated n-Lie algebroid.

Let $\left( M,\{.,...,.\}\right) $ be a Nambu-Poisson manifold. Leibniz
property for $\{.,...,.\}$ tells us that we can express it through a
n-multivector $P\in \Gamma (\Lambda ^{n}TM)$ (called the Nambu-Poisson
multivector) such that, for any $f_{1},...,f_{n}\in C^{\infty }(M)$: 
\begin{equation*}
\{f_{1},...,f_{n}\}=P_{f_{1}...f_{n}}=i_{df_{n}\wedge ...\wedge df_{1}}P.
\end{equation*}

Let us translate the Fundamental Identity for $\{.,...,.\}$ in terms of $P$.
Consider a hamiltonian vector field, which in this case will have the form 
\begin{equation*}
P_{f_{1}...f_{n-1}}\in \mathcal{X}(M)
\end{equation*}
and let us compute the Lie derivative of $P\in \Gamma (\Lambda ^{n}TM)$, $%
\mathcal{L}_{P_{f_{1}...f_{n-1}}}P$ , which is a tensor of the same type of $%
P$ and, accordingly, will act on $n$ functions $g_{1},...,g_{n}\in C^{\infty
}(M)$: 
\begin{gather*}
\left( \mathcal{L}_{P_{f_{1}...f_{n-1}}}P\right) \left(
g_{1},...,g_{n}\right) = \\
=\mathcal{L}_{P_{f_{1}...f_{n-1}}}(P_{g_{1},...,g_{n}})-\sum%
\limits_{i=1}^{n}P\left( g_{1},...,\mathcal{L}%
_{P_{f_{1}...f_{n-1}}}g_{i},...,g_{n}\right) = \\
=\mathcal{L}_{P_{f_{1}...f_{n-1}}}(\{g_{1},...,g_{n}\})-\sum%
\limits_{i=1}^{n}\{g_{1},...,\mathcal{L}%
_{P_{f_{1}...f_{n-1}}}g_{i},...,g_{n}\}= \\
=\{f_{1},...,f_{n-1},\{g_{1},...,g_{n}\}\}-\sum\limits_{i=1}^{n}\{g_{1},...,%
\{f_{1},...,f_{n-1},g_{i}\},...,g_{n}\}.
\end{gather*}

Thus, the Fundamental Identity for the n-bracket $\{.,...,.\}$, translates
into 
\begin{equation*}
\mathcal{L}_{P_{f_{1}...f_{n-1}}}P=0,\forall f_{i}\in C^{\infty }(M),i\in
\{1,...,n-1\}.
\end{equation*}

Now, this Lie derivative is nothing but the Schouten-Nijenhuis bracket of $%
P_{f_{1}...f_{n-1}}$ and $P$ (taken as multivectors), so we can write 
\begin{equation*}
\lbrack P_{f_{1}...f_{n-1}},P]_{SN}=0.
\end{equation*}

Knowing what the Fundamental Identity means in terms of $P$, we can
construct the promised example. In order to do this, consider the operator 
\begin{equation*}
D=\mathcal{L}_{P}\in \mathcal{D}_{n}^{-(n-1)}
\end{equation*}
which is of the order and degree we have seen can generate Filippov brackets
on 1-forms.

\begin{remark}
An easy computation (using that $d^{2}=0$ and $[i_{multivector},f]=0$) gives 
\begin{equation*}
\lbrack \mathcal{L}%
_{P},f]=[[i_{P},d],f]=[i_{P},[d,f]]-(-1)^{2}[d,[i_{P},f]]=[i_{P},df]=i_{i_{df}P}
\end{equation*}
and similarly (using here that $\mathcal{L}_{P}$ and $d$ commute) 
\begin{equation*}
\lbrack \mathcal{L}_{P},df]=[\mathcal{L}_{P},[d,f]]=-[d,[\mathcal{L}%
_{P},f]]=-[d,i_{i_{df}P}]=\mathcal{L}_{i_{df}P}.
\end{equation*}
\end{remark}

As a consequence we get the following result.

\begin{theorem}
For each $P\in \Gamma (\Lambda ^{n}TM)$ satisfying $%
[P_{f_{1}...f_{n}},P]_{SN}=0$ (for all $f_{i}\in C^{\infty }(M)$, $i\in
\{1,...,n-1\})$, the operator $D=\mathcal{L}_{P}\in \mathcal{D}_{n}^{-(n-1)}$
induces an n-Lie bracket on 1-forms, and so each Nambu-Poisson bracket on
functions has an associated n-Lie bracket on $\Omega ^{1}(M)$.
\end{theorem}

\begin{proof}
By Leibniz property, we only need to check that 
\begin{equation*}
\lbrack \mathcal{L}_{P},[[...[[\mathcal{L}_{P},a_{1}],a_{2}],...],a_{n-1}]]%
\in \mathcal{D}_{n-1}
\end{equation*}
when the $a_{i}^{\prime }s$ are of the form $f\in C^{\infty }(M),df\in
\Omega ^{1}(M)$ (in fact, we will see that this expression vanishes). Now,
we can distinguish three different cases.

1$^{st}$ case: there are at least two functions among the $a_{i}$ ($i\in
\{1,...,n-1\}$).

Rearrange the factors to get 
\begin{equation*}
\lbrack \mathcal{L}_{P},[[...[[[\mathcal{L}%
_{P},f_{1}],f_{2}],df_{3}],...],df_{n-1}]]
\end{equation*}
and then compute, having in mind the previous remark, 
\begin{equation*}
\lbrack \mathcal{L}_{P},[[...[[[\mathcal{L}%
_{P},f_{1}],f_{2}],df_{3}],...],df_{n-1}]]=[[...[[i_{i_{df}P},f_{2}],df_{3}],...],df_{n-1}]=0
\end{equation*}

2$^{nd}$ case: there is exactly one function among the $a_{i}$ ($i\in
\{1,...,n-1\}$).

This time, rearrange to 
\begin{gather*}
\lbrack \mathcal{L}_{P},[[[...[[\mathcal{L}%
_{P},df_{1}],df_{2}],...],df_{n-2}],f_{n-1}]]= \\
=[\mathcal{L}_{P},[[[...[\mathcal{L}%
_{_{i_{df_{1}}P}},df_{2}],...],df_{n-2}],f_{n-1}]]= \\
=...= \\
=[\mathcal{L}_{P},[\mathcal{L}_{i_{df_{n-2}\wedge ...\wedge
df_{1}}P},f_{n-1}]]=[\mathcal{L}_{P},[\mathcal{L}%
_{P_{f_{1}...f_{n-2}}},f_{n-1}]]= \\
=[\mathcal{L}_{P},i_{i_{df_{n-1}}P_{f_{1}...f_{n-2}}}]=[\mathcal{L}%
_{P},i_{P_{f_{1}...f_{n-1}}}]=i_{[P_{f_{1}...f_{n-1}},P]_{SN}}=0
\end{gather*}

3$^{rd}$ case: there is no function among the $a_{i}$ ($i\in \{1,...,n-1\}$).

Here, all we have are exact $1$-forms, and then 
\begin{gather*}
\lbrack \mathcal{L}_{P},[[...[[\mathcal{L}%
_{P},df_{1}],df_{2}],...],df_{n-1}]]= \\
=[\mathcal{L}_{P},[[...[\mathcal{L}_{_{i_{df_{1}}P}},df_{2}],...],df_{n-1}]]=
\\
=...= \\
=[\mathcal{L}_{P},\mathcal{L}_{i_{df_{n-1}\wedge ...\wedge df_{1}}P}]=[%
\mathcal{L}_{P},\mathcal{L}_{P_{f_{1}...f_{n-1}}}]=\mathcal{L}%
_{[P_{f_{1}...f_{n-1}},P]_{SN}}=0.
\end{gather*}
\end{proof}

\begin{corollary}
Let $\left( M,\{.,...,.\}\right) $ be a Nambu-Poisson manifold and $P\in
\Gamma (\Lambda ^{n}TM)$ the induced Nambu-Poisson tensor. Then $\left(
T^{\ast }M\overset{\pi }{\longrightarrow }M,[\![.,...,.]\!]_{\mathcal{L}%
_{P}},q\right) $ constructed as above is an n-Lie algebroid.
\end{corollary}

\section{Nambu-Poisson structures induced by differential operators}

We have just seen how to construct Filippov algebroids from Nambu-Poisson
structures. Next, we would like to obtain examples of the later ones also by
using differential operators techniques.

Given an operator $D\in \mathcal{D}_{n}^{-(n-1)}$ and a function $f\in
C^{\infty }(M)$, we define a bracket 
\begin{align*}
\lbrack \![.,...,.]\!]_{D_{f}}& :\Omega ^{1}(M)\times .\overset{n-1}{.}%
.\times \Omega ^{1}(M)\longrightarrow C^{\infty }(M) \\
(a_{1},...,a_{n-1})& \longmapsto \lbrack
\![a_{1},...,a_{n-1}]\!]_{D_{f}}=\Phi _{D}^{n}(a_{1},...,a_{n-1},f)
\end{align*}
and from it, a n-bracket on functions 
\begin{eqnarray*}
\{.,...,.\}_{D} &:&C^{\infty }(M)\times .\overset{n}{.}.\times C^{\infty
}(M)\longrightarrow C^{\infty }(M) \\
(f_{1},...,f_{n}) &\longmapsto
&\{f_{1},...,f_{n}\}_{D}=[\![df_{1},...,df_{n-1}]\!]_{D_{f_{n}}}=\Phi
_{D}^{n}(df_{1},...,df_{n-1},f_{n}).
\end{eqnarray*}

This bracket is linear in each argument and that it has the Leibniz
property. The following result, specifies another features.

\begin{proposition}
Let $D\in \mathcal{D}_{n}^{-(n-1)}$ and $\{.,...,.\}_{D}$ as above. Then:

(i) $\{.,...,.\}_{D}$ is skew-symmetric if and only if $Symb_{n}([D,d])=0.$

(ii) $\{.,...,.\}_{D}$ verifies the Fundamental Identity if and only if for
all $a_{1},...,a_{n-1}\in \Omega ^{1}(M),$ then $%
[D,[[...[D,a_{1}],...],a_{n-1}]]\in \mathcal{D}_{n-1}.$
\end{proposition}

\begin{proof}
Condition (ii) is known from Section 1 (see (\ref{eq5})). For the condition
(i) to be understood, we only need to check it. First, note that if $i<n-1$,
then clearly $\{f_{1},...,f_{i},f_{i+1},...,f_{n}\}=-%
\{f_{1},...,f_{i+1},f_{i},...,f_{n}\}$. Next, consider any differential
operator $\Delta $ and $f,g\in C^{\infty }(M)$; we have 
\begin{gather*}
\lbrack \lbrack \Delta ,df],g]=[[\Delta ,[d,f]],g]= \\
=[[[\Delta ,d],f],g]+(-1)^{\Delta }[d,[[\Delta ,f],g]]-[[\Delta ,f],dg]
\end{gather*}
thus, if we take $\Delta =[[...[D,df_{1}],...],df_{n-2}]\in \mathcal{D}%
_{2}^{-1}(\Omega (M)),f=f_{n-1},g=f_{n}\in C^{\infty }(M)$ it results $%
[[\Delta ,f],g]=0$, and so 
\begin{gather}
\lbrack \lbrack \Delta ,df_{n-1}],f_{n}]=  \label{eq7} \\
=[[[\Delta ,d],f_{n-1}],f_{n}]-[[\Delta ,f_{n-1}],df_{n}]=  \notag \\
=[[[\Delta ,d],f_{n-1}],f_{n}]-[[\Delta ,df_{n}],f_{n-1}].  \notag
\end{gather}
It is the first term on the last member what destroys skew-symmetry, but
writing 
\begin{equation*}
\Delta _{k-1}=[[...[D,df_{1}],...],df_{k-1}]
\end{equation*}
we see that 
\begin{eqnarray}
\lbrack \Delta _{k},d]
&=&[[[...[D,df_{1}],...],df_{k}],d]=[[[...[D,df_{1}],...],[d,f_{k}]],d]=
\label{eq8} \\
&=&[[\Delta _{k-1},[d,f_{k}]],d]=-(-1)^{\Delta _{k-1}}[df_{k},[\Delta
_{k-1},d]]=  \notag \\
&=&[[\Delta _{k-1},d],df_{k}]  \notag
\end{eqnarray}
thus, we have from (\ref{eq7}) and (\ref{eq8}), 
\begin{gather*}
\{f_{1},...,f_{n-1},f_{n}\}=[[[...[D,df_{1}],...],df_{n-1}],f_{n}](1)= \\
=[[[...[[D,d],df_{1}],...],f_{n-1}],f_{n}](1)-[[[...[D,df_{1}],...],df_{n}],f_{n-1}](1)=
\\
=[[[...[[D,d],df_{1}],...],df_{n-1}],f_{n}](1)-\{f_{1},...,f_{n},f_{n-1}\}
\end{gather*}
\end{proof}

\begin{corollary}
Let $D\in \mathcal{D}_{n}^{-(n-1)}$ be such that $%
[D,[[...[D,a_{1}],...],a_{n-1}]]\in \mathcal{D}_{n-1}$, for all $%
a_{1},...,a_{n-1}\in \Omega ^{1}(M)$, and Simb$_{n}([D,d])=0$. Then, the
induced bracket on functions $\{.,...,.\}_{D}$, is a Nambu-Poisson bracket.
\end{corollary}

Now, we prove that any Nambu-Poisson bracket comes from an operator of this
kind.

\begin{theorem}
Let $\{.,...,.\}$ be a Nambu-Poisson bracket on $C^{\infty }(M)\times .%
\overset{n}{.}.\times C^{\infty }(M)$. Then, it coincides with $%
\{.,...,.\}_{D}$, where $D=\mathcal{L}_{P}$ and $P$ is the Nambu-Poisson
n-multivector associated to $\{.,...,.\}$.
\end{theorem}

\begin{proof}
If $f_{1},...,f_{n}\in C^{\infty }(M)$, we have 
\begin{gather*}
\{f_{1},...,f_{n}\}_{\mathcal{L}_{P}}=[\![df_{1},...,df_{n-1},f_{n}]\!]_{%
\mathcal{L}_{P}}= \\
=[[[...[\mathcal{L}_{P},df_{1}],...],df_{n-1}],f_{n}](1)=\mathcal{L}%
_{i_{df_{n-1}\wedge ...\wedge df_{1}}P}f_{n}= \\
=(i_{df_{n-1}\wedge ...\wedge df_{1}}P)(df_{n})=i_{df_{n}}(i_{df_{n-1}\wedge
...\wedge df_{1}}P)=P(df_{1}\wedge ...\wedge df_{n-1}\wedge df_{n})= \\
=\{f_{1},...,f_{n}\}.
\end{gather*}
\end{proof}

Our last result shows that, in a sense, the operators of the form $D=%
\mathcal{L}_{P}$ are the unique ones for which $\{.,...,.\}_{D}$ is a
Nambu-Poisson structure. We will need some technical results first.

\begin{lemma}[of localization]
Let $p:E\longrightarrow M$ be a vector bundle with finite rank over a
manifold $M$. Let $\mathcal{E}$ \ be the space of its sections. Then, given
a linear map $A:\mathcal{E}\longrightarrow C^{\infty }(M)$, there exists a
(necessarily unique) section $\alpha $ of the dual vector bundle $E^{\ast
}\longrightarrow M$ such that, for any point $x\in M$ and any element $X$ in 
$\mathcal{E}$, 
\begin{equation*}
A(X)(x)=\alpha (X(x))
\end{equation*}
if and only if $A$ is $C^{\infty }(M)$-linear.
\end{lemma}

(This is a standard result in Differential Geometry, see for example [War
71], pgs 64-65).

\begin{definition}
A differential operator $D$ (of any order and degree) is said to be
tensorial if, for all $f\in C^{\infty }(M)$, 
\begin{equation*}
\lbrack D,f]=0.
\end{equation*}
\end{definition}

\begin{lemma}
Let $D\in \mathcal{D}_{n}^{-(n-1)}$ be a tensorial operator. Then, there
exist uniques $A\in \Gamma (\Lambda ^{n-1}TM)),\Delta \in \Gamma (\Lambda
^{n}TM\otimes T^{\ast }M)$ such that 
\begin{equation*}
D=i_{A}+i_{\Delta }.
\end{equation*}
\end{lemma}

\begin{proof}
As a consequence of $[D,f]=0$, we have that for all $\alpha \in \Omega
^{n-1}(M)$, $D(f\alpha )=fD(\alpha )$, so $D|_{\Omega ^{n-1}(M)}:\Omega
^{n-1}(M)\longrightarrow C^{\infty }(M)$ is $C^{\infty }(M)-$linear and -by
the localization lemma- it defines an $A\in \Gamma (\Lambda ^{n-1}TM))$ such
that 
\begin{equation*}
D|_{\Omega ^{n-1}(M)}=i_{A}|_{\Omega ^{n-1}(M)}.
\end{equation*}

Next, we study what happens when $D-i_{A}$ acts on $n-$forms. Just because $%
D $ and $i_{A}$ are so, $D-i_{A}$ is a tensorial operator, and then 
\begin{gather*}
\tilde{\Delta}:\Omega ^{1}(M)\times .\overset{n)}{.}.\times \Omega
^{1}(M)\times \mathcal{X}(M)\longrightarrow C^{\infty }(M) \\
(a_{1},...,a_{n},X)\longmapsto ((D-i_{A})(a_{1}\wedge ...\wedge a_{n}))(X)
\end{gather*}
is $C^{\infty }(M)-$ linear in all its arguments. Then, again by the
localization lemma, $\exists \Delta \in \Gamma (\Lambda ^{n}TM\otimes
T^{\ast }M)$ such that 
\begin{equation*}
i_{\Delta }|_{\Omega ^{n}(M)}=D-i_{A}|_{\Omega ^{n}(M)}.
\end{equation*}

Now, acting on any $k-$form, with $k<n-1$, any of the previous operators
gives $0$ (note the degrees); thus, we have $D=i_{A}+i_{\Delta }.$
\end{proof}

\begin{proposition}
Given $D\in \mathcal{D}_{n}^{-(n-1)}$, then $Symb_{n}([D,d])=0$ if and only
if $Symb_{n}(D)=Symb(\mathcal{L}_{N})$ for some $N\in \Gamma (\Lambda
^{n}TM) $.
\end{proposition}

\begin{proof}
We will follow the ideas presented in [Bel-Mon 94], where the authors
consider the $n=2$ case.

If $Symb_{n}(D)=Symb(\mathcal{L}_{N})$, where $N\in \Gamma (\Lambda ^{n}TM)$%
, we can write $D=\mathcal{L}_{N}+\tilde{D}$, with $\tilde{D}\in \mathcal{D}%
_{n-1}$, and then, as $\mathcal{L}_{N}$ commutes with $d$, 
\begin{equation*}
\lbrack D,d]=[\tilde{D},d]\in \mathcal{D}_{n-1}
\end{equation*}
so $Symb_{n}([D,d])=0.$

For the converse, consider an arbitrary $D\in \mathcal{D}_{n}^{-(n-1)}$,
neither necessarily verifying $Symb_{n}([D,d])=0$, nor being tensorial. Let
us determine its non tensorial part by taking brackets with a function $f\in
C^{\infty }(M)$. We have $[D,f]\in \mathcal{D}_{n-1}^{-(n-1)}$, so $\exists
H_{f}\in \Gamma \Lambda ^{n-1}(TM)$ such that $[D,f]=i_{H_{f}}$. Now, the
mapping 
\begin{gather*}
H:C^{\infty }(M)\rightarrow \Gamma \Lambda ^{n-1}(TM) \\
f\longmapsto H_{f}
\end{gather*}
is a derivation: if $f,g\in C^{\infty }(M)$, by the Leibniz property for the
bracket $[.,.]$, 
\begin{equation*}
i_{H_{fg}}=[D,fg]=f[D,g]+g[D,f]=fi_{H_{g}}+gi_{H_{f}}.
\end{equation*}

Now, to each derivation from $C^{\infty }(M)$ to $\Gamma \Lambda ^{n-1}(TM)$%
, we can assign a $Q\in \Gamma (TM\otimes \Lambda ^{n-1}(TM))$ in the
following manner. Let 
\begin{multline*}
\tilde{Q}:\Gamma \Lambda ^{n-1}(T^{\ast }M)\rightarrow \Gamma (TM) \\
(a_{1},...,a_{n-1})\longmapsto \tilde{Q}(a_{1},...,a_{n-1}):C^{\infty
}(M)\rightarrow C^{\infty }(M) \\
f\longmapsto i_{H_{f}}(a_{1}\wedge ...\wedge a_{n-1})=[D,f](a_{1}\wedge
...\wedge a_{n-1})
\end{multline*}
and note that $\tilde{Q}(a_{1},...,a_{n-1})\in \Gamma (TM)$ as a consequence
of $H$ being a derivation. Also, we have $i_{H_{f}}(ga_{1}\wedge ...\wedge
a_{n-1})=gi_{H_{f}}(a_{1}\wedge ...\wedge a_{n-1})$, so $\tilde{Q}%
(a_{1},...,a_{n-1})$ is $C^{\infty }(M)$- linear and there exists a $Q\in
\Gamma (TM\otimes \Lambda ^{n-1}(TM))$ such that 
\begin{equation*}
Q(df,a_{1},...,a_{n-1})=i_{\tilde{Q}(a_{1},...,a_{n-1})}df=[D,f](a_{1}\wedge
...\wedge a_{n-1}).
\end{equation*}

Let us study under what conditions this $Q$ is skew-symmetric. It suffices
to consider only the case when $a_{i}=df_{i}\in \Omega ^{1}(M),i\in
\{2,...,n\},f=f_{1}\in C^{\infty }(M)$, and to observe that 
\begin{equation*}
Q(df_{1},df_{2},...,df_{n})=[[...[[D,f_{1}],df_{2}],...],df_{n}](1)=\Phi
_{D}^{n}(f_{1},df_{2},...,df_{n}).
\end{equation*}

Now, from the proof of Proposition 10, we know that this is skew-symmetric
if and only if $Symb_{n}([D,d])=0$. So, under this condition we can take 
\begin{equation*}
N=Q\in \Gamma (\Lambda ^{n}(TM)).
\end{equation*}

In the last step, we check that, for any $f\in C^{\infty }(M)$, $[D-\mathcal{%
L}_{N},f]$ is a tensorial operator. As $[D-\mathcal{L}_{N},f]\in \mathcal{D}%
_{n-1}^{-(n-1)}$ and a differential operator is characterized by its action
on forms of degree less or equal to its order, we only need to consider the
case of a $(n-1)-$form. So, for $g_{i}\in C^{\infty }(M),i\in \{1,n-1\}$, we
compute 
\begin{gather*}
\lbrack D-\mathcal{L}_{N},f](dg_{1}\wedge ...\wedge dg_{n-1})= \\
=[D,f](dg_{1}\wedge ...\wedge dg_{n-1})-[\mathcal{L}_{N},f](dg_{1}\wedge
...\wedge dg_{n-1})= \\
=Q(df,dg_{1},...,dg_{n-1})-i_{i_{df}N}(dg_{1}\wedge ...\wedge dg_{n-1})= \\
=Q(df,dg_{1},...,dg_{n-1})-N(df,dg_{1},...,dg_{n-1})=0.
\end{gather*}

Thus, $D-\mathcal{L}_{N}$ is a tensorial operator, and applying the previous
lemma, $\exists A\in \Gamma (\Lambda ^{n-1}TM)),\Delta \in \Gamma (\Lambda
^{n}TM\otimes T^{\ast }M)$ such that 
\begin{equation*}
D=\mathcal{L}_{N}+i_{A}+i_{\Delta }
\end{equation*}
and then $[D,d]=\mathcal{L}_{A}+\mathcal{L}_{\Delta }$, with $\mathcal{L}%
_{A}\in \mathcal{D}_{n-1}^{-(n-2)}$ and $\mathcal{L}_{\Delta }\in \mathcal{D}%
_{n}^{-(n-2)}$. If $Symb_{n}([D,d])=0$, then must be $\mathcal{L}_{\Delta
}=0 $, so $\Delta =0$ and $Symb_{n}(D)=Symb(\mathcal{L}_{N})$.
\end{proof}

\begin{remark}
For $n=2$, this result appeared in \emph{[Bel 95]}.
\end{remark}

\begin{corollary}
Given $D\in \mathcal{D}_{n}^{-(n-1)}$, then $[D,d]=0$ if and only if $D=%
\mathcal{L}_{N}$ for some $N\in \Gamma (\Lambda ^{n}TM)$.
\end{corollary}

\begin{proof}
If $D=\mathcal{L}_{N}$ for some $N\in \Gamma (\Lambda ^{n}TM)$, it is clear
that $[D,d]=0$. For the converse, we have in the proof of the preceding
proposition $0=[D,d]=\mathcal{L}_{A}+\mathcal{L}_{\Delta }$, with $\mathcal{L%
}_{A}\in \mathcal{D}_{n-1}^{-(n-2)}$ and $\mathcal{L}_{\Delta }\in \mathcal{D%
}_{n}^{-(n-2)}$, so $\Delta =0$, $A=0$ and $D=\mathcal{L}_{N}$.
\end{proof}

\section{Examples}

We present here two examples of Nambu-Poisson structures with their
corresponding n-Lie algebroids. As it is not very usual to find explicit
examples in the literature, we shall be rather detailed.

\begin{enumerate}
\item  \textbf{Kepler Dynamics}. This example is based on [MVV 98]. It is
well known that the Kepler dynamics has five first integrals, which are
given by the components of the angular momentum and those of the Runge-Lenz
vector. In action-angle coordinates $J,\varphi $, such integrals are $%
J_{1},J_{2},J_{3},\varphi _{1}-\varphi _{2},\varphi _{2}-\varphi _{3}$. Call
them $h_{1},h_{2},h_{3},h_{4},h_{5}$ respectively. In the space $M$ with
coordinates $J,\varphi $, consider the Nambu-Poisson $6$-vector 
\begin{equation}
P=\frac{2mk^{2}}{(J_{1}+J_{2}+J_{3})^{3}}\frac{\partial }{\partial J_{1}}%
\wedge \frac{\partial }{\partial J_{2}}\wedge \frac{\partial }{\partial J_{3}%
}\wedge \frac{\partial }{\partial \varphi _{1}}\wedge \frac{\partial }{%
\partial \varphi _{2}}\wedge \frac{\partial }{\partial \varphi _{3}}
\label{eq9}
\end{equation}
and the induced Nambu-Poisson bracket 
\begin{equation*}
\{f_{1},f_{2},f_{3},f_{4},f_{5},f_{6}\}=P(df_{1}\wedge df_{2}\wedge
df_{3}\wedge df_{4}\wedge df_{5}\wedge df_{6}).
\end{equation*}
It is immediate to prove that the hamiltonian vector field corresponding to
the multi-hamiltonian $(h_{1},h_{2},h_{3},h_{4},h_{5})$ is 
\begin{equation*}
X_{h_{1},h_{2},h_{3},h_{4},h_{5}}=\frac{2mk^{2}}{(J_{1}+J_{2}+J_{3})^{3}}(%
\frac{\partial }{\partial \varphi _{1}}+\frac{\partial }{\partial \varphi
_{2}}+\frac{\partial }{\partial \varphi _{3}}),
\end{equation*}
so that $X_{h_{1},h_{2},h_{3},h_{4},h_{5}}(g)=%
\{h_{1},h_{2},h_{3},h_{4},h_{5},g\}$ for all $g\in C^{\infty }(M)$. Now, we
can describe the associated Filippov algebroid. The vector bundle is $%
p:\Lambda ^{1}(T^{\ast }M)\longrightarrow M$, the Filippov bracket on the
space of sections $\Omega ^{1}(M)$ is given by $\mathcal{L}_{P}$, where $P$
is the Nambu-Poisson $6$-vector (\ref{eq9}); explicitly, we would write for $%
df_{1},df_{2},df_{3},df_{4},df_{5},df_{6}\in \Omega ^{1}(M)$%
\begin{gather*}
\lbrack \!{}[df_{1},df_{2},df_{3},df_{4},df_{5},df_{6}]\!{}]_{\mathcal{L}%
_{P}}= \\
=[[[[[[\mathcal{L}_{P},df_{1}],df_{2}],df_{3}],df_{4}],df_{5}],df_{6}](1)= \\
=\mathcal{L}_{i_{df_{1}\wedge df_{2}\wedge df_{3}\wedge df_{4}\wedge
df_{5}}P}(df_{6})= \\
=d(P(df_{1}\wedge df_{2}\wedge df_{3}\wedge df_{4}\wedge df_{5}\wedge
df_{6}))= \\
=d\{f_{1},f_{2},f_{3},f_{4},f_{5},f_{6}\}.
\end{gather*}
Finally, the anchor map $q:\Omega ^{5}(M)\longrightarrow \Gamma (TM)$, acts
as 
\begin{equation*}
q(df_{1}\wedge df_{2}\wedge df_{3}\wedge df_{4}\wedge
df_{5})=[\![df_{1},df_{2},df_{3},df_{4},df_{5},.]\!]_{\mathcal{L}_{P}},
\end{equation*}
thus 
\begin{equation*}
q(df_{1}\wedge df_{2}\wedge df_{3}\wedge df_{4}\wedge
df_{5})(g)=\{f_{1},f_{2},f_{3},f_{4},f_{5},g\}.
\end{equation*}

\item  \textbf{Bihamiltonian systems}. This example is adapted from [Mag-Mag
91], and it is a system of the Calogero-Moser type. Consider two particles
on a line, interacting through a potential proportional to the inverse
square power of their distance 
\begin{equation*}
V=\frac{1}{(x_{2}-x_{1})^{2}}.
\end{equation*}

The newtonian equations of motion are readily derived: 
\begin{equation}
\left\{ 
\begin{array}{c}
\ddot{x}_{1}=-\frac{2}{(x_{2}-x_{1})^{3}} \\ 
\ddot{x}_{2}=\frac{2}{(x_{2}-x_{1})^{3}}
\end{array}
\right.  \label{eq10}
\end{equation}

These equations can also be obtained from a hamiltonian description. It
suffices to take the hamiltonian 
\begin{equation*}
H=\frac{1}{2}(p_{1}^{2}+p_{2}^{2})+\frac{1}{(x_{2}-x_{1})^{2}}
\end{equation*}
and the canonical symplectic form on the phase space $T^{\ast }\mathbb{R}%
^{2} $, with coordinates $(x_{1},x_{2},p_{1},p_{2})$: 
\begin{equation*}
\Omega =dx_{1}\wedge dp_{1}+dx_{2}\wedge dp_{2}
\end{equation*}

In fact, this system is a bihamiltonian one. We could consider, along with $%
H $, another conserved quantity: the total momentum 
\begin{equation*}
K=p_{1}+p_{2}
\end{equation*}
and the symplectic form 
\begin{eqnarray*}
\Xi &=&(p_{1}+\alpha )dx_{1}\wedge dp_{1}+(p_{2}+\alpha )dx_{2}\wedge dp_{2}+
\\
&&+\beta dx_{1}\wedge dx_{2}+\alpha (dx_{2}\wedge dp_{1}+dx_{1}\wedge
dp_{2})+\gamma dp_{1}\wedge dp_{2}
\end{eqnarray*}
where 
\begin{eqnarray*}
\alpha &=&\frac{p_{1}-p_{2}}{4+(x_{1}-x_{2})^{2}(p_{1}-p_{2})^{2}} \\
\beta &=&\frac{2(x_{1}-x_{2})}{4+(x_{1}-x_{2})^{2}(p_{1}-p_{2})^{2}} \\
\gamma &=&\frac{2}{(x_{1}-x_{3})^{3}}
\end{eqnarray*}
and we would obtain the same evolution equations (\ref{eq10}). What we are
now going to do, is to construct a Nambu-Poisson system from this
bihamiltonian one. For this purpose, it is better to introduce new
coordinates for the position of the particles: 
\begin{eqnarray*}
z &=&x_{1} \\
r &=&x_{2}-x_{1}
\end{eqnarray*}

It is straightforward to prove that the hamiltonians now adopt the form 
\begin{eqnarray*}
H &=&p_{z}^{2}+p_{r}p_{z}+\frac{1}{2}p_{r}^{2}+\frac{1}{r^{2}} \\
K &=&2p_{z}+p_{r}
\end{eqnarray*}
(the expressions for the symplectic forms are, of course, also changed; for
example now we have 
\begin{equation*}
\Omega =2dz\wedge dp_{z}+dz\wedge dp_{r}+dr\wedge dp_{z}+dr\wedge dp_{r}%
\text{).}
\end{equation*}
Consider the Nambu-Poisson multivector 
\begin{equation*}
P=\frac{\partial }{\partial r}\wedge \frac{\partial }{\partial p_{z}}\wedge 
\frac{\partial }{\partial p_{r}}\text{,}
\end{equation*}
associated to which we have a Nambu-Poisson bracket $\{.,...,.\}$, so we can
compute the hamiltonian vector field with respect to the $2$-hamiltonian $%
(H,K)$: this is the vector field 
\begin{equation*}
X_{H,K}=-p_{r}\frac{\partial }{\partial r}+\frac{2}{r^{3}}\frac{\partial }{%
\partial p_{z}}-\frac{4}{r^{3}}\frac{\partial }{\partial p_{r}}
\end{equation*}
as can be readily seen by evaluating $\{H,K,g\}=P(dH\wedge dK\wedge dg)$ for
an arbitrary function $g$. Then, it is quite easy to show the Filippov
algebroid corresponding to this structure. It is given by the vector bundle $%
p:\Lambda ^{1}(T^{\ast }M)\longrightarrow M$, and as in the previous
example, the bracket on the space of sections is determined by $\mathcal{L}%
_{P}$: 
\begin{gather*}
\lbrack \!{}[df_{1},df_{2},df_{3}]\!{}]_{\mathcal{L}_{P}}= \\
=[[[\mathcal{L}_{P},df_{1}],df_{2}],df_{3}](1)= \\
=\mathcal{L}_{i_{df_{1}\wedge df_{2}}P}(df_{3})= \\
=d(P(df_{1}\wedge df_{2}\wedge df_{3}))= \\
=d\{f_{1},f_{2},f_{3}\}.
\end{gather*}
and the anchor map $q:\Omega ^{2}(M)\longrightarrow \Gamma (TM)$, acts as 
\begin{equation*}
q(df_{1}\wedge df_{2})=[\![df_{1},df_{2},.]\!]_{\mathcal{L}_{P}},
\end{equation*}
where 
\begin{equation*}
q(df_{1}\wedge df_{2})(g)=\{f_{1},f_{2},g\}.
\end{equation*}
\end{enumerate}

\end{document}